\documentclass[12pt]{article}

\usepackage{latexsym}

\begin{document}

\title{Newman-Janis method and rotating dilaton-axion
black hole}

\author{S. Yazadjiev \thanks{E-mail: yazad@phys.uni-sofia.bg}\\
{\footnotesize  Department of Theoretical Physics,
                Faculty of Physics, Sofia University,}\\
{\footnotesize  5 James Bourchier Boulevard, Sofia~1164, Bulgaria }\\
}

\maketitle

\begin{abstract}

It's shown that the rotating dilaton-axion black hole solution can be
obtained from GGHS static charged dilaton black hole solution via
Newman-Janis method.

\noindent{PACS number(s): 02.30.Dk,04.20.Cv}
\end{abstract}
\sloppy
\renewcommand{\baselinestretch}{1.3} %
\newcommand{\sla}[1]{{\hspace{1pt}/\!\!\!\hspace{-.5pt}#1\,\,\,}\!\!}
\newcommand{\db}{\,\,{\bar {}\!\!d}\!\,\hspace{0.5pt}}
\newcommand{\lambdab}{\,\,{\bar {}\!\!\lambda}\!\,\hspace{0.5pt}}
\newcommand{\partb}{\,\,{\bar {}\!\!\!\partial}\!\,\hspace{0.5pt}}
\newcommand{\dsla}{\partb}
\newcommand{\Boxb}{\Box^{\hskip -7.7pt {}^{-}}}
\newcommand{\BoxD}{\Box^{\hskip -7.2pt {}^{{}^{D}}}}
\newcommand{\nablaD}{\nabla^{\hskip -7.2pt {}^{{}^{D}}}}
\newcommand{\GammaD}{\Gamma^{\hskip -6.2pt {}^{{}^{D}}}}
\newcommand{\SigmaD}{\Sigma^{\hskip -7.pt {}^{{}^{D}}}}
\newcommand{\ThetaD}{\Theta^{\hskip -7.pt {}^{{}^{D}}}}
\newcommand{\LambdaD}{\Lambda^{\hskip -7.pt {}^{{}^{D}}}}
\newcommand{\AD}{{\stackrel{\hskip 5.pt{}_{\it D}}{A}}}
\newcommand{\DD}{{\stackrel{\hskip 3.5pt{}_{\it D}}{D}}}
\newcommand{\ND}{{\stackrel{\hskip 2.7pt{}_{\it D}}{N}}}
\newcommand{\RD}{{\stackrel{\hskip 2.7pt{}_{\it D}}{R}}}
\newcommand{\SD}{{\stackrel{\hskip 3.7pt{}_{\it D}}{S}}}
\newcommand{\lfrac}[2]{{#1}/{#2}}
\newcommand{\sfrac}[2]{{\small \,\,\hbox{${\frac {#1} {#2}}$}}}
\newcommand{\ben}{\begin{eqnarray}}
\newcommand{\een}{\end{eqnarray}}
\newcommand{\la}{\label}
%
The low energy limit of the heterotic string theory gives an interesting
generalization of Einstein-Maxwell theory - the Einstein-Maxwell-dilaton -
axion gravity.The field equations of the Einstein-Maxwell-dilaton-axion
gravity  in four dimensions can be obtained from the following action
\cite{GSW},\cite{SS}
\ben
{\cal A}= -{1\over 16\pi} \int d^4x \sqrt{-g}
\left(R - 2\partial_{\mu}\varphi\partial^{\mu}\varphi  -
{1\over 2} e^{4\varphi}\partial_{\mu}\Theta\partial^{\mu}\Theta   +
e^{-2\varphi}F_{\mu\mu}F^{\mu\nu}  +
\Theta F_{\mu\nu}{\tilde F}^{\mu\nu}\right)
\een
Here $R$ is the Ricci scalar with respect to the space-time metric
$g_{\mu\nu}$(with a signature $(+,-,-,-)$), $\varphi$ is the dilaton field,
$F_{\mu\nu}=(dA)_{\mu\nu}$ and
${\tilde F}_{\mu\nu}$  are  correspondingly the Maxwell
tensor and its dual, the pseudo scalar $\Theta$ is related to the Kalb-Ramond
field $H^{\mu\nu\sigma}$ through the relation
$$H^{\mu\nu\sigma}={1\over 2}e^{4\varphi}\varepsilon^{\mu\nu\sigma\rho}
\partial_{\rho}\Theta.$$

In last decade the string black holes attract much
attention.The static spherically symmetric charged dilaton black  hole was
obtained by Gibbons \cite {G} and independently by Garfinkle,Horowitz and
Strominger \cite{GHS}.Using the string target space duality rotation Sen
found the rotating dilaton-axion black  hole solution generating it from
Kerr solution \cite{Sen}.

It's well-known that  Kerr and Kerr-Newman solution in Einstein theory
can be generated correspondingly from Schwarzschild and Reissner-Nordsrom
solution via Newman-Janis method \cite{NJ},\cite{NCCEPT}.
It's natural to ask whether Sen's rotating dilaton-axion solution can be
obtained via Newman-Janis method from GGHS  dilaton black hole
solution.

The purpose of the present note is to show that the rotating dilaton-axion
black hole solution  can  be "derived" from static spherically symmetric
dilaton black hole solution via Newman-Janis procedure.

Here we will not discuss the Newman-Janis algorithm in details.We refer
the reader to the recent papers \cite{DT},\cite{DS}.It should be noted,
however that in Newman-Janis procedure there is a certain arbitrariness and
an element of guess.

The GGHS dilaton black hole solution may be written in different coordinates
and there is no pure physical reasons which of them are more
appropriate for our purpose.
It seems to be natural to expect that the desirable coordinates
in which the GGHS solution should be written are these obtained by
generating the GGHS solution directly from Schwarzschild solution.
The generating the GGHS solution from Schwarzschild's one has been
already done in \cite{Y}.Here we give the final result
\ben
\la{DBH}
ds^2 =\left({1 - {r_{1}\over r}\over 1 + {r_{2}\over r}}\right)dt^2  -
\left({1 - {r_{1}\over r}\over 1 + {r_{2}\over r}}\right)^{-1} dr^2
- r^2\left(1 + {r_{2}\over r}\right)
\left(d\theta^2 + \sin^2(\theta)d\phi^2\right)   \\ \nonumber
e^{2\varphi} = {1\over 1 + {r_{2} \over r}} \\ \nonumber
\Phi = - {{Q\over r}\over 1 + {r_{2} \over r}}
\een
where $\varphi$ is the dilaton and $\Phi$ is the electric potential.
The parameters $r_{1}$  and  $r_{2}$  are given by
$$r_{1} + r_{2} =2{\cal M}$$  and $$r_{2}= {Q^2 \over {\cal M}}$$
where ${\cal M}$ and  $Q$  are the  mass and the charge of the dilaton
black hole.

Following  Newman and Janis (see also \cite{DT} and \cite{DS}) the
first step is to write the metric  (\ref{DBH}) in advanced
Eddington-Finkelstein coordinates.Performing the coordinate transformation
\ben
dt = du + \left({1 - {r_{1}\over r}\over 1 + {r_{2}\over r}}\right)^{-1}dr
\een
we obtain
\ben
\la{EF}
ds^2 = \left({1 - {r_{1}\over r}\over 1 + {r_{2}\over r}}\right)du^2
+ 2dudr - r^2\left(1 + {r_{2}\over r}\right)d\Omega^2
\een
This  metric may be presented in terms of its null tetrad vectors
\ben
g^{\mu\nu}=l^{\mu}n^{\nu} +  l^{\nu}n^{\mu}  - m^{\mu}{\bar m}^{\nu} -
m^{\nu}{\bar m}^{\mu}
\een
where
\ben
\la{TETRAD}
l^{\mu} = {\delta_{1}^{\mu}}  \\ \nonumber
n^{\mu} = {\delta_{0}^{\mu}}  -
{1\over 2}\left({1 - {r_{1}\over r}\over 1 + {r_{2}\over r}}\right)
{\delta_{1}^{\mu}}    \\ \nonumber
m^{\mu} = {1 \over \sqrt{2}r\sqrt{1 + {r_{2}\over r} }}
\left({\delta_{2}^{\mu}}   + {i\over \sin(\theta)}{\delta_{3}^{\mu}}\right)
\een
Let's now the radial coordinate $r$ allowed to take the complex  values,
as keeping the null vectors $l^{\mu}$ and $n^{\mu}$  real and
${\bar m}^{\mu}$ complex conjugated to $m^{\mu}.$
Then the tetrad takes the form
\ben
\la{COMTET}
l^{\mu} = {\delta_{1}^{\mu}}  \\ \nonumber
n^{\mu} = {\delta_{0}^{\mu}}  -
{1\over 2}
\left({1 - {r_{1}\over 2}\left({1\over r}  + {1\over {\bar r}}\right)
\over 1 + {r_{2}\over 2}\left({1\over r}  + {1\over {\bar r}}\right)}\right)
{\delta_{1}^{\mu}}    \\ \nonumber
m^{\mu} ={1 \over \sqrt{2} {\bar r}
\sqrt{1 + {r_{2}\over 2}\left({1\over r}  + {1\over {\bar r}}\right)}}
\left({\delta_{2}^{\mu}}   + {i\over \sin(\theta)}{\delta_{3}^{\mu}}\right)
\een
The next step is to perform formally the complex coordinate transformation
\ben
r^{\prime} = r + ia\cos(\theta) \,\,\,\,\,\,\,\, \theta^{{}{\prime}}=\theta
\\ \nonumber
u^{\prime} = u - ia\cos{\theta} \,\,\,\,\,\,\,\,  \phi^{\prime} =\phi
\een
By keeping $r^{\prime}$  and  $u^{\prime}$  real
we obtain the following tetrad
\ben
{l^{\prime}}^{\mu}={\delta_{1}^{\mu}}  \\ \nonumber
{n^{\prime}}^{\mu}={\delta_{0}^{\mu}}  -
{1\over 2} \left({1 - {r_{1}r^{\prime} \over \Sigma}
\over 1 + {r_{2}r^{\prime} \over \Sigma}}
\right)\delta_{1}^{\mu}     \\ \nonumber
{m^{\prime}}^{\mu}={1\over \sqrt{2}(r^{\prime} + ia\cos(\theta))}
{1\over \sqrt{1 + {r_{2}r^{\prime} \over \Sigma}}}
\left(ia\cos(\theta)(\delta_{0}^{\mu} - \delta_{1}^{\mu})  +
\delta_{2}^{\mu}  + {i\over \sin(\theta)}\delta_{3}^{\mu}\right)
\een
where $\Sigma={r^{\prime}}^2 + {a^2}\cos^{2}(\theta).$

The metric formed by this tetrad is (dropping the primes)
\ben
g^{\mu\nu}= \pmatrix{ - {a^2\sin^{2}(\theta)\over {\tilde \Sigma}} &
1 +  {a^2\sin^{2}(\theta)\over {\tilde \Sigma}}  & 0  &
- {a\over {\tilde \Sigma}}    \cr    * &
- e^{2U(r,\theta)} - {a^2\sin^{2}(\theta)\over {\tilde \Sigma}}  & 0
& {a\over {\tilde \Sigma}}  \cr
* & * & -{1\over {\tilde \Sigma}} & 0   \cr
* & * & * & - {1\over {\tilde \Sigma}} }
\een
where we have put
\ben
e^{2U(r,\theta)}= \left({1 - {r_{1}r\over \Sigma} \over
1 + {r_{2}r\over \Sigma} }\right)
\een
and
\ben
{\tilde \Sigma}= \left(1 + {r_{2}r \over \Sigma} \right) \Sigma =
r(r + r_{2}) + a^2\cos^{2}(\theta)
\een

The corresponding covariant metric is
\ben
g_{\mu\nu}= \pmatrix{ e^{2U(r,\theta)} & 1 & 0 &
a\sin^{2}(\theta)\left(1 - e^{2U(r,\theta)}\right)  \cr
* & 0 & 0 & - a\sin^{2}(\theta)   \cr
* & * & - {\tilde \Sigma} & 0    \cr
* & * & * & - sin^{2}(\theta)\left({\tilde \Sigma}  + {a^2}\sin^2(\theta)
\left(2 - e^{2U(r,\theta)}\right) \right) }
\een
A further simplification is made by the following coordinate transformation
\ben
du= dt^{\prime}  - {\Delta_{2}\over \Delta}dr \,\,\,\,\,\,\,\,
d\phi = d\phi^{\prime} - {a \over \Delta}dr
\een
where $\Delta =r(r - r_{1}) + a^2$  and $\Delta_{2} =r(r + r_{2}) + a^2.$
This transformation  leaves only one off-diagonal element and the metric
takes the form (dropping the primes on $t$ and $\phi$)
\ben
g_{\mu\nu}dx^{\mu}dx^{\nu} =
e^{2U(r,\theta)}dt^2  - {{\tilde \Sigma}
\over e^{2U(r,\theta)}{\tilde \Sigma} + {a^2}\sin^{2}(\theta)}dr^2  -
{\tilde \Sigma}d\theta^2     +  \\ \nonumber
2a\sin^{2}(\theta)\left(1 -  e^{2U(r,\theta)}\right)dtd\phi    -
\sin^{2}(\theta)\left({\tilde \Sigma} +
a^2\sin^{2}(\theta)\left(2 - e^{2U(r,\theta)}\right)\right)d\phi^2
\een
Taking into account that $r_{1} +  r_{2} = 2{\cal M} $ we obtain
\ben
\la{RDAM}
ds^2 = g_{\mu\nu}dx^{\mu}dx^{\nu} =
\left(1 - {2{\cal M}r \over {\tilde \Sigma}}\right)dt^2 -
{\tilde \Sigma}\left({dr^2 \over \Delta}  +  d\theta^2 \right) + \\ \nonumber
{4{\cal M}ra\sin^2(\theta) \over {\tilde \Sigma}}dtd\phi  -
\left(r(r + r_{2})  + a^2  +
{2{\cal M}ra^2\sin^{2}(\theta) \over {\tilde \Sigma} }\right)
\sin^2(\theta)d\phi^2
\een
where  $e^{2U(r,\theta)}{\tilde \Sigma} + a^2\sin^{2}(\theta)
= r(r - r_{1}) + a^2 = \Delta .$

This is the rotating dilaton-axion black hole metric \cite{Sen}.The other
quantities are given by
\ben
A = -{Qr \over {\tilde \Sigma}}\left(dt - a\sin^2(\theta)d\phi\right) \\
\nonumber
e^{2\varphi}={1 \over 1 + {r_{2}r \over \Sigma}}
={\Sigma \over {\tilde \Sigma}} =
{r^2 + {a^2}\cos^2(\theta) \over r(r + {Q^2 \over {\cal M}}) +
{a^2}\cos^2(\theta)}
\\ \nonumber
\Theta = {Q^2\over {\cal M}}{a\cos(\theta)\over \Sigma} =
{Q^2\over {\cal M}}{a\cos(\theta) \over r^2 + {a^2}\cos^2(\theta)}
\een
It's useful to present  the metric (\ref{RDAM}) in the form
\ben
\la{HMET}
ds^2 =e^{2U}(dt + \omega_{i}dx^{i}) -
e^{-2U}h_{ij}dx^{i}dx^{j}
\een
After a few algebra we find
\ben
ds^2 = e^{2U(r,\theta)}\left(dt +
{2{\cal M}ar\sin^2(\theta) \over {\tilde \Sigma}_{1}}d\phi\right)^2  -
\\ \nonumber
e^{-2U(r,\theta)} \left({\tilde \Sigma}_{1}
\left({dr^2 \over \Delta} + d\theta^2 \right) + \Delta\sin^2(\theta)
d\phi^2\right)
\een
where
\ben
{\tilde \Sigma}_{1}={\tilde \Sigma} - 2{\cal M}=
r(r - r_{1})   + {a^2}\cos^2(\theta) \\ \nonumber
e^{2U(r,\theta)} = 1 - {2{\cal M}r \over {\tilde \Sigma}}=
{1 \over 1 - {2{\cal M}r \over {\tilde \Sigma}_{1}}}.
\een
It should be expected that using the Newamn-Janis method
we will able  to generate stationary axisymmetric solutions
starting with static spherically symmetric solutions different from
the GGHS solution.For example, using as seed solutions the three classes
two-parametric families of solutions presented in \cite{Y}, it should be
expected that  we will obtain the corresponding rotating naked singularities
in Einstein-Maxwell-dilaton-axion gravity.

There are some questions which arise.As we have seen the Newman-Janis
method  generates the rotating solution of
Einstein-Maxwell-dilaton-axion starting with GGHS solution in proper
coordinates.GGHS solution, however, is also a solution to the truncated
theory without axion field (i.e. Einstein-Maxwell-dilaton gravity).Why
the Newman-Janis method  does not generate the rotating
solution to truncated model instead to the full model?
In our opinion the cause is  probably  that the full theory in the
presence of two commuting Killing's vectors possesses larger nontrivial
symmetry group than the truncated model.

\bigskip
\bigskip
\noindent{\Large\bf Acknowledgments}
\bigskip

The author wishes to express his thanks to P.Fiziev
for his continuous encouragement and the stimulating conversations.

This work was partially supported by the Sofia University Foundation for
Scientific Researches, Contract~No.~245/99, and by the Bulgarian National
Foundation for Scientific Researches, Contract~F610/99.

\end{document}